\documentclass[aps,prb,reprint,superscriptaddress]{revtex4-2}
\usepackage{graphicx}
\usepackage{amsmath}
\usepackage{amssymb}
\usepackage{natbib}

\begin{document}

\title{Jahn-Teller distortion driven ferromagnetism in a perovskite fluoride monolayer}
\author{Ke Ji}
\author{Zongshuo Wu}
\affiliation{School of Materials Science and Physics, China University of Mining and Technology, Xuzhou 221116, China}
\author{Xiaofan Shen}
\affiliation{National Laboratory of Solid State Microstructures and Physics School, Nanjing University, Nanjing 210093, China}
\author{Jianli Wang}
\author{Junting Zhang}
\email{juntingzhang@cumt.edu.cn}
\affiliation{School of Materials Science and Physics, China University of Mining and Technology, Xuzhou 221116, China}

\begin{abstract}
The Jahn-Teller distortion and the resulting orbital order usually cause some fascinating correlated electronic behaviors, and generally lead to antiferromagnetism in perovskite bulks. Here we demonstrate that the Jahn-Teller distortion present in the perovskite fluoride KCrF$_3$ bulk can be retained to the two-dimensional limit, resulting in a staggered orbital order and ferromagnetism in the perovskite monolayer. Octahedral tilt and rotation distortion also appear in the ground-state structure of the perovskite monolayer, which have minor effects on the electronic and magnetic properties with respect to the Jahn-Teller distortion. In addition, in the prototype phase without structural distortion, the partial occupation of the $e_g$ orbitals leads to a ferromagnetic metallic state. This work facilitates the design of two-dimensional ferromagnets and functional properties based on Jahn-Teller distortion and orbital order.
\end{abstract}
\maketitle

\section{Introduction}
Two-dimensional (2D) ferromagnetism has attracted considerable interest due to the potential applications in next-generation spintronics devices with high density, high processing speed, and low energy consumption \cite{Wang2018,Song2018,Klein2018,Kim2018,Gong2019,Jiang2021}. However, the existence of 2D ferromagnetism was once a long-standing open question because the long-range magnetic order is prohibited at finite temperature in the 2D isotropic Heisenberg model according to the Mermin-Wagner theorem \cite{Mermin1966}. Until 2017, 2D ferromagnetism was first experimentally confirmed in the van der Waals materials CrI$_3$ and Cr$_2$Ge$_2$Te$_6$ \cite{Gong2017,Huang2017}. Subsequently, many 2D ferromagnets have been discovered experimentally or proposed theoretically, however mostly focused on van der Waals materials \cite{Bonilla2018,Deng2018,Zhu2018,Gibertini2019,Huang2018,Wang2020}.

Compared with van der Waals materials, perovskite materials exhibit more abundant correlated electronic behaviors, such as giant magnetoresistance, superconductivity, and magnetoelectric multiferroics, due to the competition and coupling between various degrees of freedom in lattice, charge, orbital and spin \cite{Dagotto2001,Dagotto2005,Eerenstein2006,Zhang2022}. However, unlike van der Waals materials that retain their structural and chemical bonding properties when reduced to a monolayer, 2D perovskites may undergo structural reconstruction with respect to their bulk phases due to the changes in ion coordination and chemical bonds. Therefore, whether some interesting phenomena found in perovskite bulks can exist in 2D perovskites remains a key issue to be solved. Recently, some stress-free perovskite monolayers have been experimentally prepared and can be transferred to any desired substrate \cite{Lu2016,Hong2017,Ji2019}. These experimental advances not only provide great opportunities for theoretical design of functional properties based on 2D perovskites such as ferroelectricity, ferromagnetism and  multiferroics \cite{Lu2017a,Zhang2020,Zhou2021,Zhou2022,Shen2023}, but also for the experimental realization of perovskite moir\'{e} superlattice via rotational misalignment and heterobilayers with different lattice constants \cite{Jia2020,Ricciardulli2021}.

Jahn-Teller (JT) distortion, as a common type of lattice distortion in perovskites, is usually associated with the breaking of the electron degeneracy of the transition-metal ions \cite{Carpenter2009}. In general, the JT distortion is accompanied by the appearance of orbital order. The most typical example is the perovskite rare-earth manganite, as a representative material of the giant magnetoresistance effect \cite{Dagotto2001}. The cooperative JT distortion leads to an in-plane staggered orbital order, which is responsible for its rich magnetic phases \cite{Kimura2003,Zhang2018,Zhang2018a}. Another typical example is  perovskite halide containing JT-active ions, such as Cr$^{2+}$ or Cu$^{2+}$ ions. The staggered orbital order tends to form an in-plane ferromagnetic order, whereas the interplane antiferromagnetic coupling causes these materials to exhibit antiferromagnetism, \emph{i.e.}, \emph{A}-type antiferromagnetic order \cite{Binggeli2004,Margadonna2006,Xiao2010,Wang2011}. It can be expected that 2D ferromagnetism may occur in the monolayers of these perovskites, if the JT distortion and the staggered orbital order can be maintained to the monolayer limit.

In this paper, we focus on a perovskite fluoride monolayer with complete octahedra (formula K$_2$CrF$_4$), and study its lattice dynamics, structural, electronic and magnetic properties through first-principles calculations. Our results show that only the JT distortion is a dynamically unstable mode, whereas the ground-state structure exhibits octahedral tilt and rotation distortion in addition to the JT distortion. The cooperative JT distortion leads to a staggered orbital order and opens an indirect band gap in the ground-state phase. Ferromagnetism is dominated by strong ferromagnetic coupling of the nearest-neighbor (NN) exchange interaction, and can exist even in the prototype phase without JT distortion and orbital order.

\section{Computational details}
The first-principles calculations based on density functional theory (DFT) were performed using the projector-augmented wave (PAW) method \cite{Bloechl1994}, as implemented in the Vienna \emph{ab initio} simulation package (VASP) \cite{Kresse1996}. The Perdew-Burke-Ernzerhof functional modified for solids (PBEsol) \cite{Perdew2009} was used as the exchange-correlation functional. Electron correlation was treated using the Hubbard-\emph{U} method within the rotationally invariant formalism \cite{Dudarev1998,Kulik2006}, and an effective value of $U_{eff} = 1.5$ eV for the Cr 3\emph{d} states was determined by self-consistent calculation based on the constrained random-phase approximate method \cite{Sasioglu2011}. Due to the periodic condition, a vacuum space of 20 {\AA} was utilized to keep adjacent images apart in the vertical direction. A plane-wave cutoff energy of 600 eV was used for the plane wave expansion, and a $\varGamma$-centered $9\times9\times1$ \emph{k}-point mesh was used for the Brillouin zone integration. The electronic self-consistent loop was performed at the convergence threshold of $10^{-6}$ eV. The in-plane lattice constants and internal atomic coordinates were relaxed until the Hellman-Feynman force on each atom was less than $10^{-2}$ eV/{\AA}.

The ISOTROPY tool \cite{Campbell2006} was used to aid with the group theoretical analysis. Phonon band structures were calculated using density functional perturbation theory (DFPT). The phonon frequencies and corresponding eigenmodes were calculated on the basis of the extracted force-constant matrices, as implemented in the PHONOPY code \cite{Togo2015}. The Nos\'{e} heat bath scheme was employed to carry out the first-principles molecular dynamics simulations in a canonical ensemble. A large supercell of $3\times3\times1$  was used to minimize the constraint of periodic boundary conditions. The magnetic exchange interaction parameters were calculated using the Green's function method with the local rigid spin rotation treated as a perturbation, as implemented in the TB2J package \cite{He2021}. The Monte Carlo simulations were performed to verify the magnetic ground state and estimate the magnetic transition temperature. A $32\times32$ supercell with periodic boundary condition and the annealing algorithm were used in simulations.

\section{results and discussion}

\subsection{Lattice dynamics}
\begin{figure}
\centering
\includegraphics*[width=0.45\textwidth]{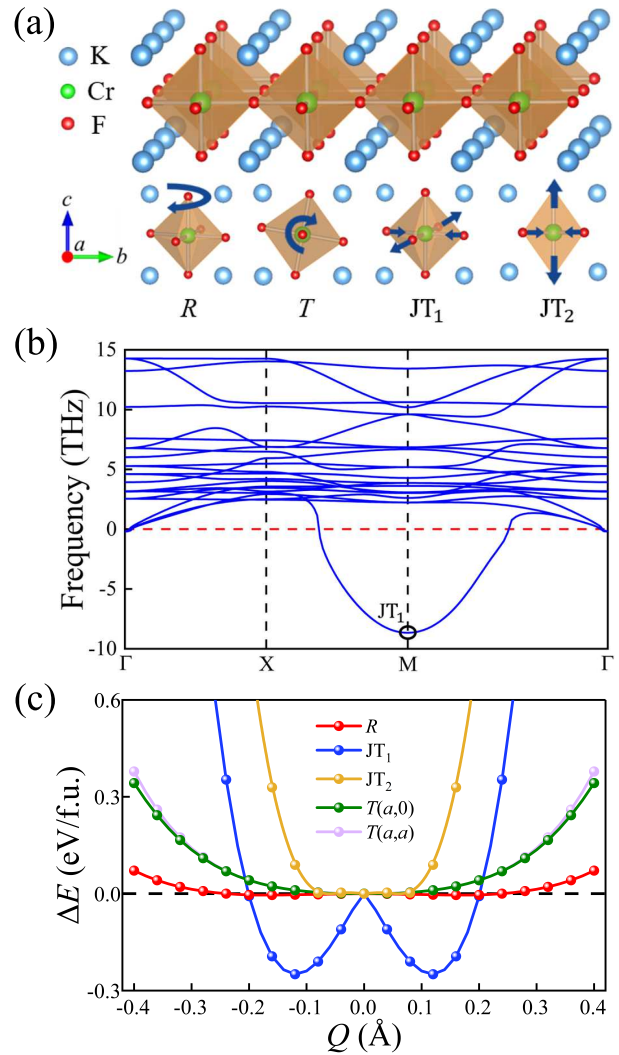}
\caption{\label{Fig1} (a) Crystal structure of the prototype phase of the perovskite monolayer and schematic diagram of the four octahedral distortion modes: rotation (\emph{R}), tilt (\emph{T}), JT$_1$, and JT$_2$. Curved arrows represent the direction of rotation of the octahedron. Straight arrows represent the direction of anion displacement due to JT distortion. (b) Phonon spectrum of the prototype phase of the K$_2$CrF$_4$ monolayer. The lowest-frequency soft mode belongs to the JT$_1$ mode. (c) Change in energy as a function of the amplitude of each distortion mode for the prototype phase. Two special in-plane directions are considered for the rotation axis of the tilt mode.}
\end{figure}

Considering the possible structural reconstruction when reduced to 2D, we start from the prototype phase of the perovskite monolayer (space group $P4/mmm$). Besides JT distortion, octahedral rotation is another common type of structural distortion that may occur in 2D perovskites. Octahedral rotation distortion is caused by cationic radius mismatch and is therefore very common in perovskites \cite{Carpenter2009}. In 2D perovskites, it can be divided into a rotation mode belonging to irreducible representation $M_2^+$ and a tilt mode belonging to irreducible representation $M_5^+$, corresponding to the rotation of the octahedron around the out-of-plane and in-plane axes, respectively, as shown in Fig. 1(a). The JT distortion can be divided into JT$_1$ and JT$_2$ modes, corresponding to the irreducible representations $M_3^+$ and $M_4^+$, respectively. The JT$_1$ mode represents the elongation of two in-plane bonds and the shortening of the other two in-plane bonds in a single octahedron, while the JT$_2$ mode refers to the elongation of the two out-of-plane bonds and the contraction of the four in-plane bonds, as shown in Fig. 1(a).

We first calculated the phonon spectrum of the prototype phase, whose in-plane lattice constant is 4.13 {\AA} after structural optimization. Figure 1(b) shows that an unstable vibrational mode (\emph{i.e.}, soft mode) occurs at the high symmetry \emph{M} point in the Brillouin zone, which belongs to the JT$_1$ mode. Then we calculated the energy gain caused by the freezing of each octahedral distortion mode. For the tilt mode with 2D irreducible representation, two different order parameters $M_5^+(a, 0)$ and $M_5^+(a, a)$ were considered, corresponding to the octahedral rotation axis along [100] and [110] directions, respectively. The results show that both the tilt mode and the JT$_2$ mode show parabolic energy curves, \emph{i.e.}, no energy gain, while the rotation mode has a slight energy gain, as shown in Fig. 1(c). In contrast, the JT$_1$ mode shows a typical double-well energy curve, that is, its emergence can significantly reduce the energy of the system. These results imply the presence of JT$_1$ distortion in the ground-state structure.

\subsection{Ground-state structure}

\begin{table}
\centering
\includegraphics*[width=0.48\textwidth]{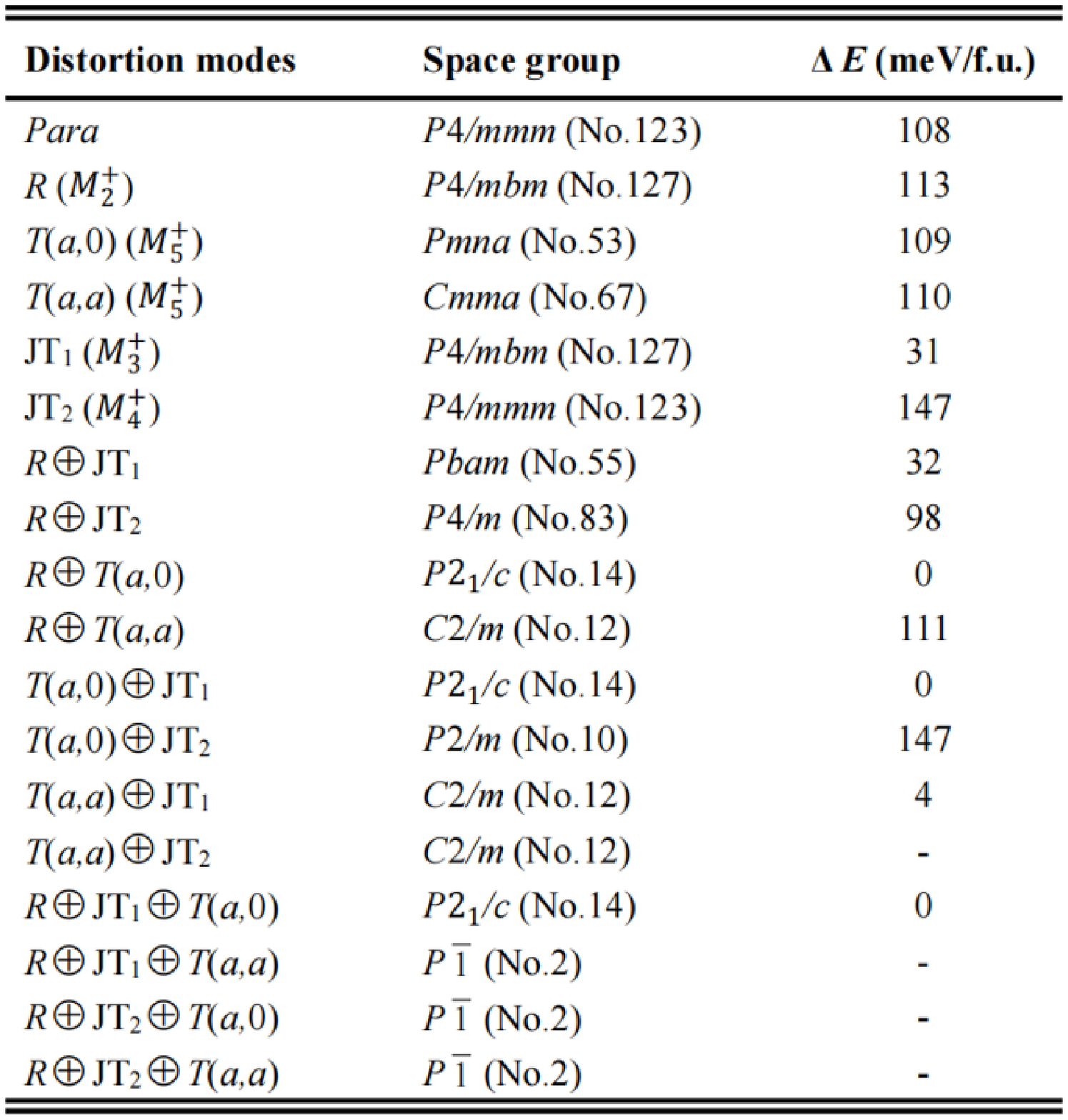}
\caption{\label{table 1} Symmetry and energy of structural phases resulting from single distortion mode and their various combinations in the K$_2$CrF$_4$ monolayer. The symbol "-" indicates that the corresponding structural phase is unstable, since it transforms into other structure phase after structural optimization.}
\end{table}

In order to determine the ground-state structure of the K$_2$CrF$_4$ monolayer, we considered the structural phases caused by each distortion mode and their various combinations (see Table \uppercase\expandafter{\romannumeral1}). We analyzed the symmetry of all structural phases, and calculated their energy by optimizing the atomic coordinates and the in-plane lattice constants under fixed structural symmetry. Of all the structural phases caused by a single distortion mode, only the structural phase of the JT$_1$ mode has lower energy than the prototype phase. For some structural phases established by multiple distortion modes, some distortion modes such as JT$_2$ will disappear after structural optimization. Furthermore, some additional distortion modes allowed by symmetry may be introduced after structural optimization, such as in the lowest-energy $P2_1/c$ phase. This ground-state phase can be established by a combination of tilt mode with rotation or JT$_1$ mode. However, all three distortion modes appear in the ground-state structure, where the JT$_1$ and tilt mode have considerable amplitude while the rotation distortion is very slight. The octahedral distortion of the ground-state structure is similar to that of the bulk phase, which exhibits both the JT$_1$ and tilt distortion at low temperature \cite{Margadonna2006,Xiao2010}. In order to verify the stability of the ground-state structure, we further calculated its phonon spectrum [see Fig. 2(a)] and performed first-principles molecular dynamics simulations [see Fig. 2(b)], which confirm its dynamic stability and thermodynamic stability, respectively.

\begin{figure}
\centering
\includegraphics*[width=0.45\textwidth]{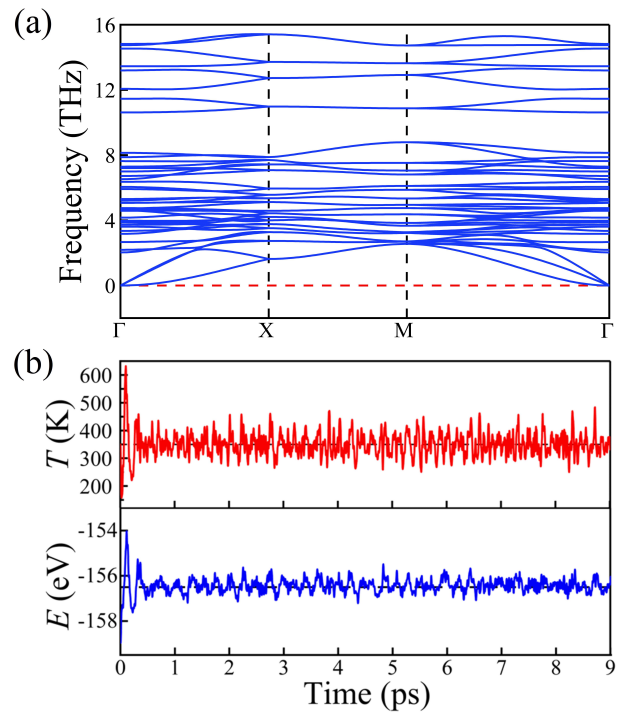}
\caption{\label{Fig2}  (a) Phonon spectrum and (b) first-principles molecular dynamics simulations at 350 K of the ground-state phase.}
\end{figure}

\subsection{Electronic property}
In order to elucidate the effect of structural distortion on the electronic property of the K$_2$CrF$_4$ monolayer, we first calculated the projected band structure of the prototype phase [see Fig. 3(a)]. In 2D perovskite, the octahedral crystal field ($O_h$ symmetry) transforms into the tetrahedral crystal field ($D_{4h}$ symmetry), which causes the double-degenerate $e_g$ orbitals to split into a lower energy $d_{x^2-y^2}$ orbital and a higher energy $d_{3z^2-r^2}$ orbital, and the triple-degenerate $t_{2g}$ orbitals to split into the single $d_{xy}$ orbital and the double-degenerate $d_{xz}/d_{yz}$ orbitals. The electronic configuration of the Cr$^{2+}$ ion is a high-spin state $t_{2g}^3e_g^1$, \emph{i.e.}, the spin-up $t_{2g}$ orbitals are fully occupied and the $e_g$ orbitals are half-full occupied. In the prototype phase, the band formed by the $d_{x^2-y^2}$ orbital shows significant dispersion due to the large orbital overlap, as shown in Fig. 3(a). The two bands formed by the $e_g$ orbitals have very close energy near the high symmetry \emph{M} point, which causes the Fermi level to cross these two bands, namely exhibiting a metallic state.

\begin{figure}
\centering
\includegraphics*[width=0.45\textwidth]{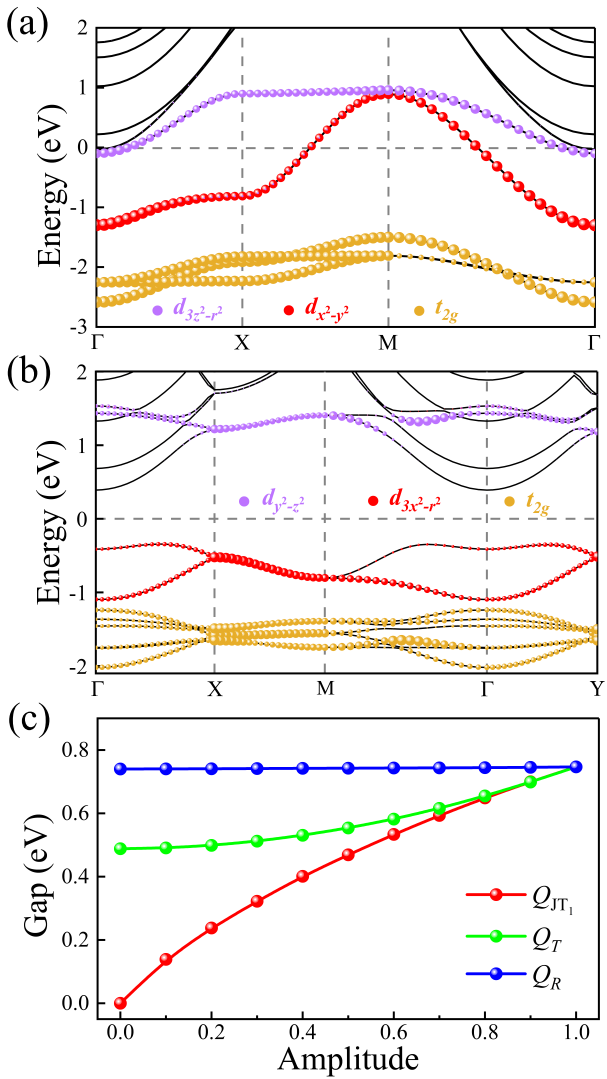}
\caption{\label{Fig3}  The up-spin projected band structures of (a) the prototype and (b) the ground-state phases. The local \emph{z} axis is along the out-of-plane direction, and the \emph{x} axis is along the longest Cr-F bond. (c) Variation of band gap with the relative amplitude of each distortion mode in the ground-state phase.}
\end{figure}

In the ground-state phase, the cooperative JT$_1$ distortion causes the longest Cr-F bond to be staggered in the \emph{ab} plane, which reduces the energy of the in-plane $e_g$ orbitals ($d_{3x^2-r^2}$ and $d_{3y^2-r^2}$). In addition, the octahedral tilt distortion causes the Cr-F-Cr bond angle to deviate significantly from 180$^\circ$. These structural distortions reduce the overlap of the in-plane $e_g$ orbitals, resulting in the suppression of band dispersion, as shown in Fig. 3(b). The bands formed by the two $e_g$ orbitals are completely separated in energy, \emph{i.e.}, an indirect band gap of about 0.73 eV is opened. We further investigated the effect of various distortion modes on the electronic property of the ground-state phase by reducing the amplitude of one distortion mode while keeping the amplitudes of other modes constant. The band gap basically does not change with the weakening of the rotation mode due to its small amplitude, but reduces significantly with that of the tilt mode, as shown in Fig. 3(c). The band gap still exists even if the tilt distortion is completely removed. However, the band gap decreases more rapidly with the weakening of the JT$_1$ mode, and disappears when the JT$_1$ mode is completely removed. Therefore, the emergence of JT$_1$ distortion plays a key role in the insulating behavior of the ground-state phase, similar to the perovskite rare-earth manganites \cite{Kimura2003}.

\subsection{Magnetic property}
Next we determined the magnetic ground state of the prototype phase. Magnetic structures with different periodic modulation are represented by the wave vector \textbf{q}, whose general value represents the spiral spin order. For high symmetry points, $\textbf{q}=\varGamma$ corresponds to the ferromagnetic order, while $\textbf{q}=M, X$ represent the collinear N\'{e}el and stripe antiferromagnetic orders, respectively. Figure 4(a) shows the variation of energy with the wave vector. The results show that the ferromagnetic order has the lowest energy, so the prototype phase exhibits a ferromagnetic metallic state. The origin of ferromagnetism of the prototype phase can be attributed to the double-exchange mechanism, since the partial occupation of the two $e_g$ orbitals resulting in the emergence of the itinerant electrons.

\begin{figure}
\centering
\includegraphics*[width=0.45\textwidth]{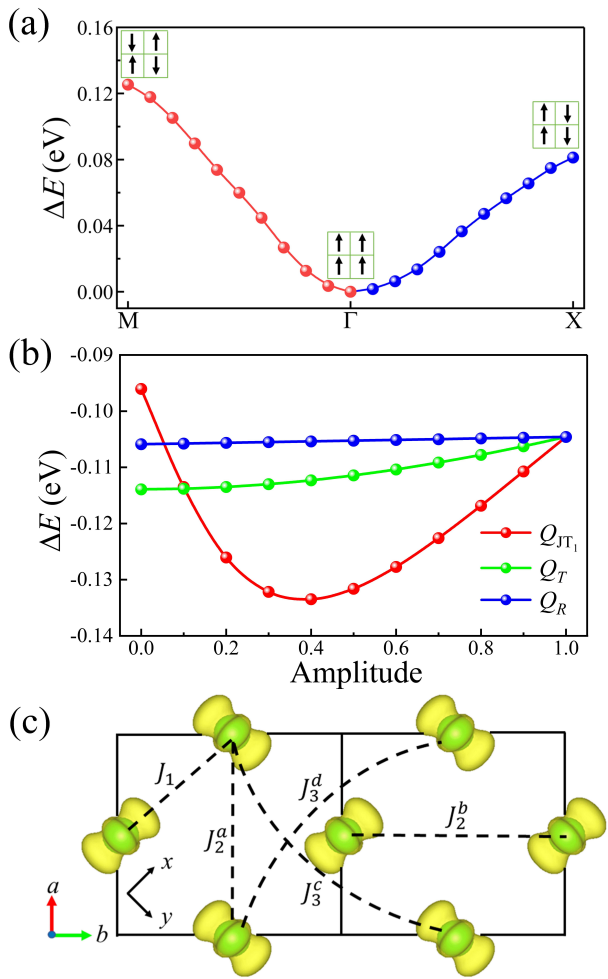}
\caption{\label{Fig4} (a) Energy of the magnetic ordering as a function of the spin spiral wave vector in the prototype phase. (b) Variation of energy difference between the ferromagnetic (FM) and the N\'{e}el antiferromagnetic (AFM) orders ($\Delta E= E_{FM}-E_{AFM}$) with the relative amplitude of each distortion mode in the ground-state phase. (c) The partial charge density at the valence band maximum and the considered exchange interactions in the ground-state phase.}
\end{figure}

For the ground-state structure, the energy of the ferromagnetic order is still lower than that of the antiferromagnetic orders. To reveal the role of different distortion modes in the emergence of ferromagnetism, we varied the amplitudes of the three distortion modes in the ground-state phase, and then calculated the variation of the energy difference between the ferromagnetic and the N\'{e}el antiferromagnetic orders with the amplitude of each mode. This energy difference actually represents the NN exchange interaction parameter. As shown in Fig. 4(b), the energy difference does not vary with the amplitude of the rotation mode, while the weakening of the tilt mode leads to a slight decrease in the energy difference. In contrast, the energy difference varies significantly and non-monotonically with the amplitude of the JT$_1$ mode. The trend of ferromagnetism is first enhanced and then weakened with the increasing amplitude of the JT$_1$ mode. This non-monotonic change may result from the competition between the two interactions. Specifically, the JT$_1$ distortion induces the occurrence of the staggered orbital order [see Fig.4(c)] that causes ferromagnetic superexchange interaction, however, its further enhancement will reduce the overlap of the occupied $d_{3x^2-r^2}/d_{3y^2-r^2}$ orbitals and thus weaken the superexchange interactions. The ferromagnetic order consistently has a lower energy throughout the varying range, indicating the robustness of the ferromagnetism.

Then we calculated the Heisenberg magnetic exchange interaction parameters by using the Green's function method based on magnetic force theorem. The NN, second-nearest-neighbor (SNN), and third-nearest-neighbor (TNN) exchange interactions are shown in Fig. 4(c). The SNN exchange interaction is split into two classes, $J_2^a$ and $J_2^b$, due to the presence of the tilt distortion. Similarly, the orbital order causes the TNN exchange interaction to split into $J_3^c$ and $J_3^d$. The NN exchange interaction results in a strong ferromagnetic coupling $J_1 = 11.75$ meV (with normalized spin moment), while the SNN exchange interaction is very weak, with values of $J_2^a = 0.36$ meV and $J_2^b = 0.22$ meV. The TNN exchange interaction shows significant anisotropy, with a considerable antiferromagnetic coupling along the $\sigma$-bond direction ($J_3^c = -1.47$ meV), but very slight along the $\pi$-bond direction ($J_3^d  = 0.02$ meV). For comparison, we recalculated the exchange interaction parameters using the four-state method \cite{Xiang2011}, with $J_1 = 12.65$ meV, $J_2^a = 0.81$ meV and $J_2^b = 0.72$ meV, $J_3^c = -1.21$ meV and $J_3^d = -0.62$ meV. The exchange interaction parameters calculated by both methods are quite close except for the very weak exchange interactions. Therefore, the ferromagnetism of the ground-state phase is dominated by the NN ferromagnetic exchange interaction, which can be explained by the Goodenough-Kanamori-Anderson (GKA) rule in superexchange interaction. According to the GKA rule, the staggered orbital order leads to a ferromagnetic $e_g$-$e_g$ exchange coupling, in addition to the ferromagnetic $e_g$-$t_{2g}$ coupling. These ferromagnetic couplings are much stronger than the antiferromagnetic coupling between the $t_{2g}$-$t_{2g}$ orbitals. These analyses are confirmed by the orbital decomposition of the magnetic exchange interaction. The ferromagnetic coupling of the NN exchange interaction mainly arises from the $d_{3x^2-r^2}$-$d_{xy}$ and $d_{3x^2-r^2}$-$d_{3y^2-r^2}$ coupling, with values of 8.49 and 2.64 meV, respectively. Therefore, the orbital order plays an important role in the emergence of ferromagnetism of the ground-state phase.

\begin{figure}
\centering
\includegraphics*[width=0.45\textwidth]{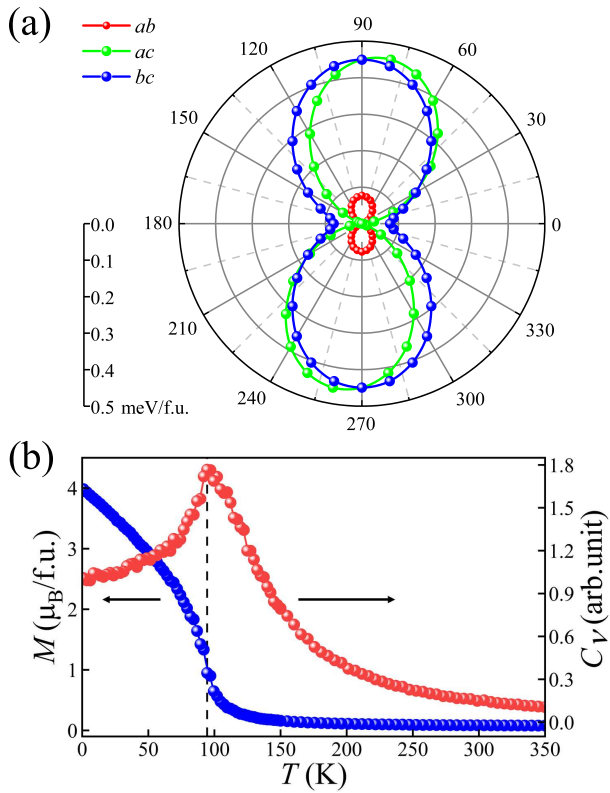}
\caption{\label{Fig5} (a) Magnetic anisotropy energy of the ground-state phase. The change in energy is shown as a function of the spin angle. When the spins lie in the \emph{ab} (\emph{ac}) and \emph{bc} planes, the spin angle refers to the \emph{a} and \emph{b} axes, respectively. (b) Monte Carlo simulations of the ground-state phase. Magnetization (\emph{M}) and specific heat ($C_v$) are shown as functions of temperature.}
\end{figure}

Magnetic anisotropy calculation shows that the in-plane spin orientation has lower energy, while the tilt distortion causes a slight in-plane anisotropy, leading to the easy-magnetization axis along the tilt axis, as shown in Fig. 5(a). The magnetic anisotropy basically originates from the spin-orbit coupling effect of Cr ion. The orbital magnetic moment of all ions is almost zero, while the spin magnetic moment is mainly concentrated on Cr ion. Then Monte Carlo simulations were performed to confirm the ferromagnetism and estimate the Curie temperature of the ground-state phase. The Heisenberg spin model containing the magnetic anisotropy energy term
$$H = -\sum_{ij}J_{ij}\textbf{S}_i\cdot\textbf{S}_j-\sum_{i}A(\textbf{S}^z_i)^2$$
was used, where $J_{ij}$ contains the NN, SNN and TNN exchange parameters, \emph{A} is the single-ion magnetic anisotropy constant, $S_i^z$ represents the component of the spin along the easy-magnetization axis. As shown in Fig. 5(b), the ferromagnetism of the ground-state phase is further confirmed by the magnetization curve, and the Curie temperature is estimated to be 94 K according to the specific curve, slightly higher than the magnetic phase transition temperature of the bulk phase \cite{Xiao2010}.
\section{conclusion}
In conclusion, we have systematically studied the lattice dynamics, structural, electronic and magnetic properties of the perovskite monolayer K$_2$CrF$_4$. Three lattice distortion modes appear in the ground-state structure: JT$_1$ distortion, octahedral tilt and rotation distortion, of which the former has a significant effect on electronic and magnetic properties. The JT$_1$ distortion causes a staggered orbital order, resulting in a ferromagnetic semiconductor for the ground-state phase. Both the ground-state and the prototype phases exhibit ferromagnetism, which is derived from the superexchange and double-exchange mechanisms, respectively. This work demonstrates that perovskite monolayers containing JT-active magnetic ions may provide a broad platform for designing 2D ferromagnets.

\begin{acknowledgments}
This work was supported by the Fundamental Research Funds for the Central Universities (Grant No. 2019QNA30). Computer resources provided by the High Performance Computing Center of Nanjing University are greatfully acknowledged.
\end{acknowledgments}

\end{document}